\documentclass[showpacs,aps,graphicx,twocolumn]{revtex4}%,preprint

\usepackage{graphicx}

\begin{document}

\title{ Genuine tripartite entanglement in quantum brachistochrone evolution of a
three-qubit system\footnote{Published in Phys. Rev. A \textbf{80},
052106 (2009)}}
\author{Bao-Kui Zhao,$^{1,2,3}$ Fu-Guo Deng,$^{4}$\footnote{
Email: fgdeng@bnu.edu.cn}  Feng-Shou Zhang,$^{1,2,3}$  and Hong-Yu
Zhou$^{1,2,3}$}
\address{$^1$Key Laboratory of Beam Technology and Material
Modification of Ministry of Education, Beijing Normal University,
Beijing 100875,  China\\
$^2$College of Nuclear Science and Technology, Beijing Normal
University,
Beijing 100875, China\\
$^3$Beijing Radiation Center, Beijing 100875,  China\\
$^4$Department of Physics, Beijing Normal University, Beijing
100875, China }
\date{\today }

\begin{abstract}

We explore the connection between quantum brachistochrone
(time-optimal) evolution of a three-qubit system and its residual
entanglement called three-tangle. The result shows that the
entanglement between two qubits is  not required for some
brachistochrone evolutions of a three-qubit system. However, the
evolution between two distinct states cannot be implemented without
its three-tangle, except for the trivial cases in which less than
three qubits attend evolution. Although both the probability density
function of the time-averaged three-tangle and that of the
time-averaged squared concurrence between two subsystems become more
and more uniform with the decrease in angles of separation between
an initial state and a final state, the features of their most
probable values exhibit a different trend.
\end{abstract}
 \maketitle

\section{introduction}

The speed of the evolution of a quantum system governed by a given
energy constraint is very important in quantum information as it
provides a useful tool for people to estimate the fundamental limits
that basic physical laws impose on how fast information can be
processed or transmitted \cite{Margolus,Levitin,Zielinski,Kosinski}.
Also, one can exploit the limit on the speed of quantum evolution to
construct an optimal quantum gate with which the quantum time
required for a system to evolve between two mutually orthogonal
states is the shortest one \cite{bgate} . However, two elementary
relations limit the speed of quantum evolution. On the one hand, the
time-energy uncertainty relation \cite{Anandan} imposes a lower
limit on the time interval $T$ taken by a quantum system to evolve
from a given state to its orthogonal one. This bound on $T$ is
related to the spread of energy of the system $\triangle E$, i.e.,
$T\geq \frac{h}{4\triangle E}$. On the other hand, Margolus and
Levitin \cite{Margolus} showed that such a quantity is also related
to the fixed average energy of the system $E$, i.e., $T\geq
\frac{h}{4E}$. These two relations together establish the  limit
time of quantum evolution speed, i.e., the minimum time
$T(E,\triangle E)$ required for a system with the energy $E$ and the
energy spread $\triangle E$ to evolve through two orthogonal states.
Recently, Giovannetti, Lloyd and Maccone
\cite{Giovannetti,Giovannettipra,GiovannettiJOB} showed that there
is an interesting connection between quantum entanglement and the
evolution speed of quantum systems. Some groups
\cite{Giovannetti,Giovannettipra,GiovannettiJOB,Kupferman} have
studied the connection between entanglement and the speed of
evolution in mixed states. This connection has been extensively
studied for the evolution between orthogonal and nonorthogonal, pure
and mixed, as well as bipartite and multipartite cases
\cite{add1,add2,add3,add4,BorrasZander,Borras}.

Quantum  brachistochrone evolution problem is used to deal with the
Hamiltonian generating the optimal quantum evolution $\vert
\psi(t)\rangle$ between two prescribed states $\vert \psi_I\rangle$
and $\vert \psi_F\rangle$ \cite{qbepprl}. That is, it searches the
shortest time interval $T$ taken by a quantum system for evolving
from an original state to a final one. This problem attracts a lot
of attention \cite{qbepprl,qbep2,CarliniHosoya}. For instance,
Carlini \emph{et al}. \cite{qbepprl} presented a general framework
for finding the time-optimal evolution and the optimal Hamiltonian
for quantum system with a given set of initial and final states.
Brody and Hook \cite{qbep2} established an elementary derivation of
the optimum Hamiltonian, under constraints on its eigenvalues, that
generates the unitary transformation $\vert \psi_I\rangle
\rightarrow \vert \psi_F\rangle$ in the shortest duration.  Carlini
\emph{et al}. \cite{CarliniHosoya} investigated quantum
brachistochrone evolution of mixed states.

Recently, Borras \emph{et al}. \cite{BorrasZander} investigated the
role of entanglement in quantum  brachistochrone evolution of
quantum system in a pure state and found that " brachistochrone
quantum evolution between orthogonal states cannot be implemented
without entanglement". Moreover, they \cite{Borras} discussed
quantum brachistochrone evolution of systems of two identical
particles and found that entanglement plays a fundamental role in
the brachistochrone evolution of composite quantum systems. That is,
quantum  brachistochrone evolution of a composite quantum systems
composed of distinguishable subsystems cannot be implemented without
entanglement if there are at least two subsystems attending
evolutions.  In these two works \cite{BorrasZander,Borras}, they
gave out the connection between the entanglement of two subsystems
(i.e., the linear entropy of two subsystems \cite{BorrasZander}  or
their squared concurrence C$^2$  \cite{Borras}) and time-optimal
quantum evolutions in the cases of two-qubit systems, two-qutrit
systems and three-qubit systems.

In three-qubit systems, there is another entanglement shared by all
the three qubits, i.e., the so-called residual entanglement by
Coffman, Kundu, and Wootters \cite{CKW}. It is termed as
three-tangle $\tau_{ABC}$ \cite{threetangle} and can be expressed as
\cite{CKW}
\begin{eqnarray}
\tau_{ABC}=C^2_{A(BC)} - C^2_{AB} - C^2_{AC},
\end{eqnarray}
where $C_{A(BC)}$ is the concurrence of the two subsystems $A$ and
$BC$. In detail,  concurrence is a useful tool for quantifying the
entanglement of bipartite quantum systems and it is given by
\cite{wotters}
\begin{eqnarray}
C_{AB}=max\{0,\lambda_1-\lambda_2-\lambda_3-\lambda_4\},
\end{eqnarray}
where $\lambda_1\geq \lambda_2\geq \lambda_3\geq \lambda_4$ are the
square roots of the eigenvalues of the matrix
$\rho_{AB}(\sigma^A_y\otimes
\sigma^B_y)\rho^*_{AB}(\sigma^A_y\otimes \sigma^B_y)$. Here
$\rho_{AB}$ is the matrix of the bipartite quantum system $AB$ and
$\rho^*_{AB}$ is its complex conjugation. $\sigma^A_y$ is the Pauli
matrix expressed in the same basis as
\begin{eqnarray}
\sigma_{y}=\left(\begin{array}{cc}
0  & -i \\
i & 0
\end{array}\right).
\end{eqnarray}
Different from concurrence $C_{A(BC)}$ which is shared only by two
subsystems $A$ and $BC$, $\tau_{ABC}$ is real entanglement shared
among the three particles of the system. That is, it is invariant
under permutations of the three qubits, i.e.,
$\tau_{ABC}=\tau_{BCA}=\tau_{CAB}$.

In this paper, we will explore the connection between three-tangle
$\tau_{ABC}$ and quantum  brachistochrone evolution of a three-qubit
system. Our result shows that in some special cases of quantum
brachistochrone evolution of a three-qubit system, the entanglement
between two subsystems $C^2_{AB}$ ($C^2_{AC}$) is not required.
However, the evolution  between two distinct states cannot be
implemented without three-tangle $\tau$, except for the trivial
cases in which there are less than three qubits attending quantum
 brachistochrone evolution. Our result is tighter than that in
Ref.\cite{BorrasZander} for quantum  brachistochrone evolution of a
three-qubit system. Moreover, we find that both the probability
density function of the time-averaged three-tangle and that of the
time-averaged squared concurrence between two subsystems become more
and more uniform with the decrease in angles of separation between
an initial state and a final state, but the features of their most
probable values exhibit a different trend.

The paper is organized as follows. In Sec. II A, we describe the way
for the calculation of time averaged three-tangle in quantum
brachistochrone evolution. In Sec. II B, we discuss the role of
three-tangle  $\tau_{ABC}$ in quantum brachistochrone evolution and
find that brachistochrone evolutions cannot be implemented without
$\tau_{ABC}$. However, the entanglement of each two qubits is not
necessary in some brachistochrone evolutions. In Secs. III A and III
B, we give the probability density of $\langle\tau\rangle$ in
quantum
 brachistochrone evolutions between two symmetric states with the
different angles of separation ($\theta/2=\pi/8,\pi/4,3\pi/8,\pi/2$)
and that between two general states, respectively. A brief
discussion and summary are given in Sec. IV.

\section{time averaged entanglement during  brachistochrone evolution}

\subsection{Time averaged three-tangle in quantum  brachistochrone evolution}

Quantum  brachistochrone evolution problem means searching the
shortest possible time for generating the optimal quantum evolution
$\vert \psi(t)\rangle$ from an initial quantum state
$|\Psi_{I}\rangle$ to a final quantum state $|\Psi_{F}\rangle$ under
the constraint that the difference between the maximum eigenenergy
and the minimum one of the Hamiltonians $H_T$ generating the unitary
transformation
$|\Psi_{I}\rangle\rightarrow|\Psi_{F}\rangle=e^{\frac{iH_T}{\hbar}}|\Psi_{I}\rangle$
is  not more than a given constant energy 2$\omega$. This constraint
is necessary as it is easy to implement a quantum evolution
connecting the two alluded states and taking an arbitrarily small
time $T$ if the differences between the eigenenergies of the
Hamiltonians are arbitrarily large \cite{BorrasZander}. The
time-optimal  evolution is given by
\cite{qbepprl,qbep2,BorrasZander}
\begin{eqnarray}
|\Psi(t)\rangle=[\text{cos}(\frac{\omega t}{\hbar})
-\frac{\text{cos}\frac{1}{2}\theta}{\text{sin}\frac{1}{2}\theta}\text{sin}(\frac{\omega t}{\hbar})]|\Psi_{I}\rangle\nonumber\\
+\frac{1}{\text{sin}\frac{1}{2}\theta}\text{sin}(\frac{\omega
t}{\hbar})|\Psi_{F}\rangle.
\end{eqnarray}
Here $|\psi(0)\rangle=|\Psi_{I}\rangle$ and
$|\psi(T)\rangle=|\Psi_{F}\rangle$ represent the initial state and
the final one, respectively, and
\begin{eqnarray}
T=\frac{\hbar\theta}{2\omega}.
\end{eqnarray}
$\theta$ can be regarded as the angle of separation of the initial
and the final states. If this pair of states are orthogonal, i.e.,
$\langle \Psi_I|\Psi_F\rangle=0$ ($\theta=\pi$), then
\begin{eqnarray}
T &=& \frac{\pi\hbar}{2\omega},\\
|\Psi(t)\rangle&=& \text{cos}(\frac{\omega
t}{\hbar})|\Psi_{I}\rangle
+\text{sin}(\frac{\omega t}{\hbar})|\Psi_{F}\rangle.\label{qbe}%\nonumber\\
\end{eqnarray}

During the time-optimal  evolution, the time averaged three-tangle
can be calculated as follows
\begin{eqnarray}
\langle\tau(\Psi(t))\rangle &=&
\frac{1}{T}\int_{0}^{T}\tau(\Psi(t))dt,
\end{eqnarray}
where $\tau(\Psi(t))$ is the three-tangle of a three-qubit system in
the state $\Psi_{ABC}(t)$. For an arbitrary three-qubit pure state
\begin{eqnarray}
|\Psi\rangle_{ABC}=\sum a_{ijk}|ijk\rangle_{ABC},\;\;\;\; (i,j,k \in
\{0,1\}),
\end{eqnarray}
its three-tangle $\tau_{ABC}(\Psi)$ can be calculated as follows:
\cite{CKW}
\begin{eqnarray}
\tau_{ABC}(\Psi)=4|d_{1}-2d_{2}+4d_{3}|,
\end{eqnarray}
where
\begin{eqnarray}
d_{1}&=& a_{000}^{2}a_{111}^{2}+a_{001}^{2}a_{110}^{2}+a_{010}^{2}a_{101}^{2}+a_{100}^{2}a_{011}^{2},\nonumber\\
d_{2}&=& a_{000}a_{111}a_{100}a_{011}+a_{000}a_{111}a_{101}a_{010}\nonumber\\
&+&a_{000}a_{111}a_{110}a_{001}+a_{011}a_{100}a_{101}a_{010}\nonumber\\
&+&a_{011}a_{100}a_{110}a_{001}+a_{010}a_{101}a_{110}a_{001},\nonumber\\
d_{3}&=&a_{000}a_{110}a_{101}a_{011}+a_{111}a_{001}a_{010}a_{100}.
\end{eqnarray}
For example, the three-tangle  $\tau$ in a standard
Greenberger-Horne-Zeilinger (GHZ) state
$|GHZ\rangle=\frac{1}{\sqrt{2}}(|000\rangle + |111\rangle)$ is 1 and
that in a W state $|W\rangle=\frac{1}{\sqrt{3}}(|001\rangle +
|010\rangle + |100\rangle)$ is 0.

\subsection{Role of three-tangle in quantum  brachistochrone evolution}

Borras \emph{et al}. \cite{BorrasZander,Borras} showed that quantum
 brachistochrone evolution between orthogonal states cannot be
implemented without the typical entanglement of two subsystems (such
as  linear entropy or concurrence). For a three-qubit entangled
quantum system, its concurrence satisfies the following relation,
\begin{eqnarray}
C^2_{A(BC)}=  C^2_{AB} + C^2_{AC} + \tau_{ABC}.
\end{eqnarray}
That is, $C^2_{A(BC)}$ includes three parts. Our question is, is
each term of the three parts in $C^2_{A(BC)}$ necessary in a quantum
brachistochrone evolution?

Let us consider two specific cases of time-optimal evolution of
three-qubit symmetric states with the following two pairs of initial
and final states to analyze the role of the three terms $ C^2_{AB}$,
$C^2_{AC}$, and $\tau_{ABC}$.
%\begin{widetext}
%\begin{center}
\begin{eqnarray}
(i) \;\; |\Psi^1_I\rangle &=&
\frac{1}{\sqrt{3}}(|110\rangle+|101\rangle+|011\rangle)\rightarrow
|\Psi^1_F\rangle\nonumber\\
 &=& \frac{\text{cos}\alpha}{\sqrt{2}}(|000\rangle
+|111\rangle)+\frac{\text{sin}\alpha}{\sqrt{3}}(|001\rangle\nonumber\\
&+& |010\rangle + |100\rangle),\label{initial1}\\%\nonumber
(ii) \;\;|\Psi^2_I\rangle &=&
\frac{1}{\sqrt{2}}(|000\rangle-i|111\rangle)\rightarrow
|\Psi^2_F\rangle\nonumber\\
&=& \frac{1}{\sqrt{2}}(i|000\rangle-|111\rangle).\label{initial2}%
\end{eqnarray}
%\end{center}
%\end{widetext}
Here $\langle \Psi^j_I|\Psi^j_F\rangle=0$ ($j=1,2$).

In the case ($i$), the three-tangle $\tau_{ABC}$ in the initial
state $|\Psi^1_I\rangle$ is 0. The final state $|\Psi^1_F\rangle$ is
a superposition of $|GHZ\rangle$ and $|W\rangle$ and its
three-tangle is determined by the coefficient $\alpha$. When
$\alpha=0$ the final state is a $|GHZ\rangle$ state and its
three-tangle is 1. When $\alpha=\frac{\pi}{2}$ the final state is a
$|W\rangle$ state and its three-tangle is 0. During the time-optimal
evolution, the state of the three-qubit system $ABC$ at the time $t$
is described by Eq.(\ref{qbe}). Let $\xi=\frac{\omega t}{\hbar}$,
the state can be written as
\begin{eqnarray}
|\Psi(\xi,\alpha)\rangle_{ABC} &=& \text{cos}\xi\frac{1}{\sqrt{3}}(|110\rangle+|101\rangle+|011\rangle)_{ABC}\nonumber\\
&+& \text{sin}\xi[\frac{\text{cos}\alpha}{\sqrt{2}}(|000\rangle+|111\rangle)_{ABC}\nonumber\\
&
+&\frac{\text{sin}\alpha}{\sqrt{3}}(|100\rangle+|010\rangle+|001\rangle)_{ABC}].\nonumber\\
\end{eqnarray}
The three-tangle in the state $|\Psi(\xi,\alpha)\rangle_{ABC}$ is
given by
\begin{eqnarray}
\tau(\xi,\alpha)_{ABC} &=& 4|\frac{1}{4}\text{sin}^{4}\xi \text{cos}^{4}\alpha
-\frac{1}{3}\text{sin}^{2}\xi \text{cos}^{2}\xi \text{sin}^{2}\alpha\nonumber\\
&& -\text{sin}^{3}\xi \text{cos}\xi \text{sin}\alpha \text{cos}^{2}\alpha \nonumber\\
&&+\frac{4}{3\sqrt{6}}\text{sin}\xi \text{cos}^{3}\xi \text{cos}\alpha\nonumber\\
&& +\frac{4}{3\sqrt{6}}\text{sin}^{4}\xi \text{sin}^{3}\alpha
cos\alpha|.
\end{eqnarray}
The time average three-tangle can be calculated by
\begin{eqnarray}
\langle\tau(\alpha)\rangle_{ABC} &=&
\frac{2}{\pi}\int_{0}^{\frac{\pi}{2}}\tau(\xi,\alpha)_{ABC}d\xi.
\end{eqnarray}

\begin{figure}[!h]%[tpb]
\begin{center}
\includegraphics[width=8cm,angle=0]{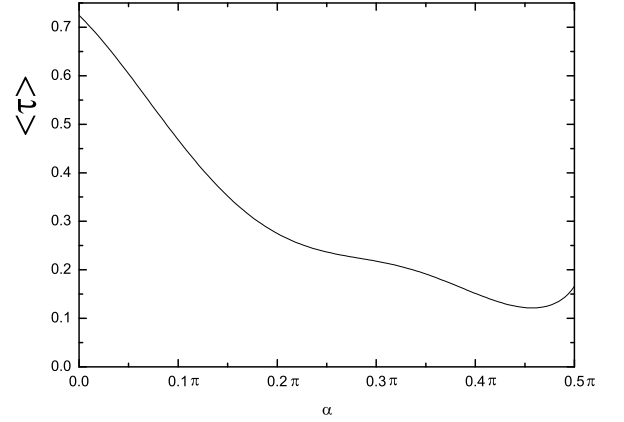}
\caption{Plot of $\langle\tau(\alpha)\rangle$ as a function of
$\alpha$ ($\alpha\in[0,\frac{\pi}{2}]$) for the time-optimal
evolutions from $|\Psi_I\rangle_1$ to
$|\Psi_F\rangle_1$.}\label{fig1}
\end{center}
\end{figure}

The relation between the time averaged three-tangle
$\langle\tau(\alpha)\rangle$ and the parameter $\alpha$ is shown in
Fig.1. For
$|\Psi^1_I\rangle=|\widetilde{W}\rangle=\frac{1}{\sqrt{3}}(|110\rangle+|101\rangle+|011\rangle)\rightarrow
|\Psi^1_F\rangle=|GHZ\rangle=\frac{1}{\sqrt{2}}(|000\rangle
+|111\rangle)$ ($\alpha=0$), $\langle\tau(0)\rangle=0.7215$; for
$|\Psi^1_I\rangle=\frac{1}{\sqrt{3}}(|110\rangle+|101\rangle+|011\rangle)\rightarrow
|\Psi^1_F\rangle=\frac{1}{\sqrt{3}}(|001\rangle+|010\rangle
+|100\rangle)$ ($\alpha=\pi/2$), $\langle\tau(\pi/2)\rangle=0.1667$.
From Fig.\ref{fig1}, one can see that  the time averaged
three-tangle is maximal when the three-qubit quantum system evolves
from the state
$|\widetilde{W}\rangle=\frac{1}{\sqrt{3}}(|110\rangle+|101\rangle+|011\rangle)$
to the state $|GHZ\rangle=\frac{1}{\sqrt{2}}(|000\rangle
+|111\rangle)$. Moreover, the time averaged three-tangle is larger
than zero, which means the three-tangle $\tau_{ABC}=1$ is necessary
in these evolutions.

In the case ($ii$), the state of the three-qubit system at the time
$t$ is
\begin{eqnarray}
|\Psi(t)\rangle= \frac{1}{\sqrt{2}}(e^{i\frac{\omega
t}{\hbar}}|000\rangle - ie^{-i\frac{\omega t}{\hbar}}|111\rangle).
\label{qbe2}
\end{eqnarray}
The three-tangle of the three-qubit system during the time-optimal
evolution is $\tau_{ABC}=C^2_{A(BC)}=1$, i.e.,
$C^2_{AB}=C^2_{AC}=0$, which means the entanglement of each two
qubits $C_{AB}$ ($C_{AC}$ or $C_{BC}$) is not required in this
time-optimal evolution. However, it requires the three-tangle
$\tau_{ABC}$. In other words, this time-optimal evolution cannot be
implemented without the three-tangle $\tau_{ABC}$.

Certainly, there is at least a class of initial states and final
states making the three-tangle be $\tau_{ABC}=C^2_{A(BC)}=1$ during
the time-optimal evolution. That is, the initial state is
\begin{eqnarray}
|\Psi^{2'}_I\rangle= \frac{1}{\sqrt{2}}(e^{i\phi_A}|klm\rangle -
e^{i\phi_B}|\bar{k}\bar{l}\bar{m}\rangle),
\end{eqnarray}
the final state is
\begin{eqnarray}
|\Psi^{2'}_F\rangle =\frac{i}{\sqrt{2}}(e^{i\phi_A}|klm\rangle +
e^{i\phi_B}|\bar{k}\bar{l}\bar{m}\rangle),
\end{eqnarray}
and the state at the time $t$ during the time-optimal evolution is
\begin{eqnarray}
|\Psi'(t)\rangle= \frac{1}{\sqrt{2}}(e^{i\frac{\omega t}{\hbar} +
i\phi_A}|klm\rangle - e^{-i\frac{\omega t}{\hbar}
+i\phi_B}|\bar{k}\bar{l}\bar{m}\rangle). \label{qbe2}
\end{eqnarray}
Here $\phi_A$ and $\phi_B$ are two arbitrary real numbers. $k,l,m
\in\{0,1\}$ and $\bar{k}=1-k$.

From these two special cases of time-optimal evolutions of
three-qubit symmetric states, one can see that the entanglement of
each two qubits $C_{AB}$ ($C_{AC}$ or $C_{BC}$) is not necessary in
some brachistochrone evolutions.  However, the  brachistochrone
evolutions cannot be implemented without the three-tangle $\tau$.

\section{three-tangle in quantum  brachistochrone evolution}

\subsection{Three-tangle in quantum  brachistochrone evolution between two symmetric states}

In order to explore the typical features of $\langle\tau\rangle$ in
all possible  brachistochrone evolutions of a three-qubit system
$ABC$ between a pair of symmetric states, we are going to sample
systematically the aforementioned set of time-optimal evolutions by
generating randomly pairs of symmetric states
$|\Psi_{I}\rangle_{ABC}$ and $|\Psi_{F}\rangle_{ABC}$ with a given
overlap $\langle\Psi_{I}|\Psi_{F}\rangle=cos(\theta/2)$, i.e.,
\begin{eqnarray}
|\Psi_{I}\rangle_{ABC}=c_{1}|000\rangle+c_{2}\frac{1}{\sqrt{3}}\{|001\rangle+|010\rangle+|100\rangle\}\nonumber\\
+c_{3}\frac{1}{\sqrt{3}}\{|110\rangle+|101\rangle+|011\rangle\}+c_{4}|111\rangle,\\%\nonumber
|\Psi_{F}\rangle_{ABC}=d_{1}|000\rangle+d_{2}\frac{1}{\sqrt{3}}\{|001\rangle+|010\rangle+|100\rangle\}\nonumber\\
+d_{3}\frac{1}{\sqrt{3}}\{|110\rangle+|101\rangle+|011\rangle\}+d_{4}|111\rangle,
\end{eqnarray}
where
\begin{eqnarray}
\sum_{k=1}^4|c_{k}^{2}| &=& \sum_{k}^4|d_{k}^{2}|=1,\\
\sum_{k=1}^4c_{k}d_{k}^{\ast} &=& cos(\theta/2).
\end{eqnarray}
It is easy to see that the three qubits $A$, $B$, and $C$ are
symmetric (uniform) in the initial state $|\Psi_{I}\rangle_{ABC}$
and the final state $|\Psi_{F}\rangle_{ABC}$.

We use the same way as Refs.\cite{BorrasZander,Borras} to generate
randomly these pairs of states. In detail, let us use
$|\Psi\rangle=\sum_{k=1}^{4}c_{k}|k\rangle$ to describe a general
state of a three-qubit system. Here $\{|k\rangle\}$ ($k=1,2,3,4$) is
a set of basis states for the three-qubit system $ABC$, i.e,
$\{|k\rangle\}=\{|000\rangle,
\frac{1}{\sqrt{3}}\{|001\rangle+|010\rangle+|100\rangle\},
\frac{1}{\sqrt{3}}\{|110\rangle+|101\rangle+|011\rangle\},
|111\rangle\}$, and $\sum_{k=1}^{4}|c_{k}|^{2}=1$. Each pair of
states $|\Psi_{I}\rangle$ and $|\Psi_{F}\rangle$ can be generated
with the Haar measure \cite{Zyczkowski} by random $4\times 4$
unitary matrices $M_{4\times 4}$ uniformly distributed upon the
vectors $\vert \Psi_{I0}\rangle=(1,0,0,0)^\dag $ and $\vert
\Psi_{F0}\rangle=(cos\frac{\theta}{2},sin\frac{\theta}{2},0,0)^\dag
$, i.e., $\vert \Psi_{I}\rangle=M_{4\times 4}\vert \Psi_{I0}\rangle$
and  $\vert \Psi_{F}\rangle=M_{4\times 4}\vert \Psi_{F0}\rangle$.
For each rotation matrix $M_{4\times 4}$,  the pair of symmetric
states $|\Psi_{I}\rangle_{ABC}$ and $|\Psi_{F}\rangle_{ABC}$ have an
overlap $\langle\Psi_{I}|\Psi_{F}\rangle=cos(\theta/2)$. As the
matrix $M_{4\times 4}$ is a random unitary one, the initial state
$\vert \Psi_I\rangle$ and the final state $\vert \Psi_F\rangle$
generated by this matrix are a random pair with an overlap
$\langle\Psi_{I}|\Psi_{F}\rangle=cos(\theta/2)$. For each of these
pairs of states, we calculate the time averaged three-tangle
$\langle\tau\rangle$ in the quantum
 brachistochrone evolution connecting the initial state
$|\Psi_{I}\rangle_{ABC}$ and the final state
$|\Psi_{F}\rangle_{ABC}$.

\begin{figure}[!h]%[tpb]
\begin{center}
\includegraphics[width=8cm,angle=0]{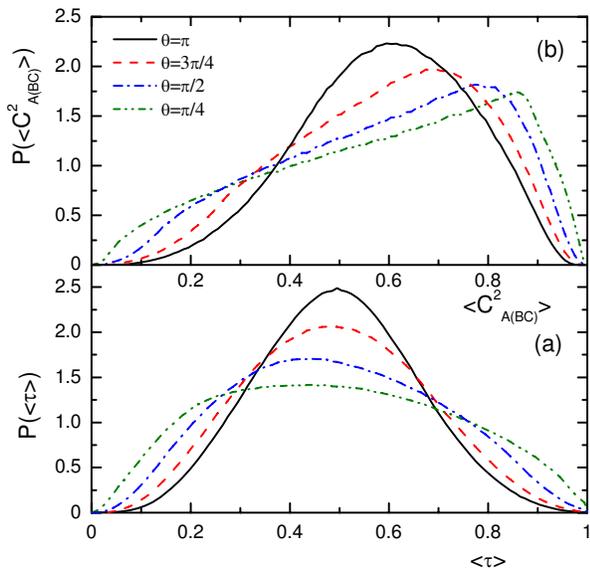}
\caption{(Color online) (a) Probability density functions
$P(\langle\tau\rangle)$ of the three-tangle $\langle\tau\rangle$
corresponding to quantum brachistochrone evolutions of a three-qubit
system between two symmetric states with different angles of
separation ($\theta/2=\pi/8,\pi/4,3\pi/8,\pi/2$); (b) probability
density functions $P(\langle C^2_{A(BC)}\rangle)$ of the time
averaged squared concurrence $\langle C^2_{A(BC)}\rangle$
corresponding to the same evolutions.}\label{fig2}
\end{center}
\end{figure}

The probability densities of $\langle\tau\rangle$ and  $P(\langle
C^2_{A(BC)}\rangle)$ in quantum
 brachistochrone evolutions between two symmetric states with the
different angles of separation ($\theta/2=\pi/8,\pi/4,3\pi/8,\pi/2$)
are shown in Fig.\ref{fig2}(a) and Fig.\ref{fig2}(b), respectively.
From this figure, one can see that both $P(\langle\tau\rangle)$ and
$P(\langle C^2_{A(BC)}\rangle)$ become more uniform when the angle
becomes smaller. The smaller the angle $\theta/2$, the larger the
most probable values of the time averaged entanglement $\langle
C^2_{A(BC)}\rangle$. On the contrary, the smaller the angle
$\theta/2$, the smaller  the most probable values of the time
averaged entanglement $\langle\tau\rangle$. Except for the trivial
evolution between $|\Psi_I\rangle$ and
$|\Psi_F\rangle=|\Psi_I\rangle$, the time-optimal evolution between
two symmetric states cannot be implemented without the three-tangle
$\tau$.

\subsection{Three-tangle in quantum  brachistochrone evolution between two general states}

\begin{figure}[!h]%[tpb]
\begin{center}
\includegraphics[width=8cm,angle=0]{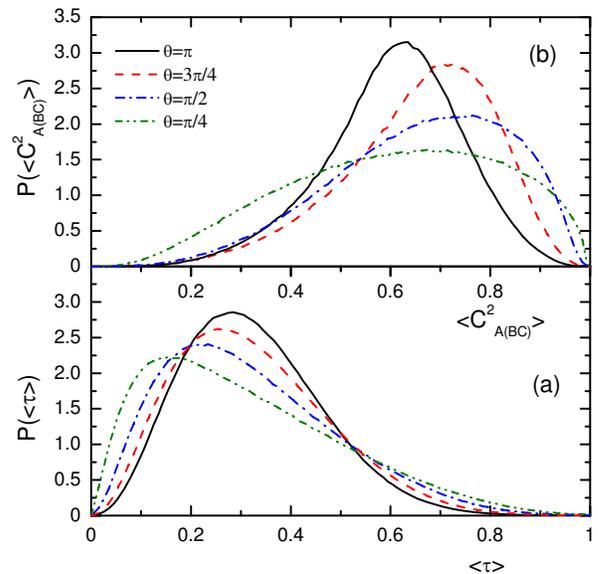}
\caption{(Color online) (a) Probability density functions
$P(\langle\tau\rangle)$ of the three-tangle $\langle\tau\rangle$
corresponding to quantum  brachistochrone evolutions of a
three-qubit system between two general states with different angles
of separation ($\theta/2=\pi/8,\pi/4,3\pi/8,\pi/2$); (b) probability
density functions $P(\langle C^2_{A(BC)}\rangle)$ of the time
averaged squared  concurrence $\langle C^2_{A(BC)}\rangle$
corresponding to the same evolutions.}\label{fig3}
\end{center}
\end{figure}

Two general states with a given overlap $cos(\theta/2)$ for a
three-qubit system can be described as follows:
\begin{eqnarray}
|\Psi_{I}\rangle &=& c_{1}|000\rangle+c_{2}|001\rangle+c_{3}|010\rangle+c_{4}|100\rangle\nonumber\\
&+& c_{5}|110\rangle+c_{6}|101\rangle+c_{7}|011\rangle\}+c_{8}|111\rangle,\\
|\Psi_{F}\rangle &=& d_{1}|000\rangle+d_{2}|001\rangle+d_{3}|010\rangle+d_{4}|100\rangle\nonumber\\
&+&
d_{5}|110\rangle+d_{6}|101\rangle+d_{7}|011\rangle+d_{8}|111\rangle,
\end{eqnarray}
where
\begin{eqnarray}
\sum_{k}^8|c_{k}|^{2} &=& \sum_{k}^8|d_{k}|^{2}=1,\\
\sum_{k}^8c_{k}d_{k}^{\ast} &=& cos(\theta/2).
\end{eqnarray}
Similar to the case with quantum  brachistochrone evolution between
two symmetric states, we can also calculate the probability density
functions $P(\langle\tau\rangle)$ of the time averaged three-tangle
$\langle\tau\rangle$ and the probability density functions
$P(\langle C^2_{A(BC)}\rangle)$ of the time averaged squared
concurrence $\langle C^2_{A(BC)}\rangle$  corresponding to the same
evolutions for $\theta/2=\pi/8,\pi/4,3\pi/8,\pi/2$, shown in
Figs.\ref{fig3}(a) and \ref{fig3}(b), respectively. From
Fig.\ref{fig3}, one can get the similar result for the case with two
symmetric states in this time.

Compared with the case between two symmetric states, the most
probable values of the time averaged entanglement
$\langle\tau\rangle$ is smaller with the same angle of separation.

\section{discussion and summary}

There are two classes of trivial evolutions in a three-qubit system.
One is the evolution between two same states and the other is that
for less than three subsystems. In the first case, the three-qubit
system does not evolute as $\vert \Psi_I\rangle=\vert
\Psi(t)\rangle=\vert \Psi_F\rangle$. In the second case, the initial
state is $\vert \Psi_I\rangle=\vert \phi\rangle_i\otimes \vert
\psi\rangle_{jk}$ and the final state is $\vert \Psi_F\rangle=\vert
\phi\rangle_i\otimes \vert \psi'\rangle_{jk}$ (here $i\neq j \neq k
\in \{A,B,C\}$). That is, the qubit $i$ is a stationary particle and
it is not correlated with the qubits $j$ and $k$ in the evolution.
Similar to the case for the system composed of two identical
particles discussed in Ref.\cite{Borras}, the time-averaged
three-tangle required for these trivial  quantum
 brachistochrone evolutions need not be greater than zero. As these
evolutions are not genuine three-qubit quantum  brachistochrone
evolutions, they do not affect our result.

In summary, we have explored the connection between three-tangle and
quantum  brachistochrone evolution of a three-qubit system. We have
shown that the evolution between two distinct states cannot be
implemented without three-tangle, except for the trivial cases in
which there are less than three qubits attending in quantum
brachistochrone evolution or the final state and the initial state
are the same one. However, the entanglement between two qubits is
not required in some quantum brachistochrone evolutions. Moreover,
we have found that both the probability density function of the
time-averaged three-tangle $\langle\tau\rangle$ and that of the
time-averaged squared concurrence $\langle C^2_{A(BC)}\rangle$
between two subsystems become more and more uniform with the
decrease in angles of separation $\theta/2$ between an initial state
and a final state. However, the features of their most probable
values exhibit a different trend. The result between two symmetric
states agrees with that between two general states.

\section*{ACKNOWLEDGMENTS}

This work is supported by the National Natural Science Foundation of
China under Grant Nos. 10604008 and 10974020, A Foundation for the
Author of National Excellent Doctoral Dissertation of P. R. China
under Grant No. 200723, and  Beijing Natural Science Foundation
under Grant No. 1082008.

\end{document}